\documentclass[reprint,amsmath,amssymb,aps,prb]{revtex4-1}

\usepackage{graphicx}
\usepackage{dcolumn}% Align table columns on decimal point
\usepackage{bm}
\usepackage{amsmath}

\begin{document}

\preprint{APS/123-QED}

\title{Enhanced piezoelectric response of AlN via CrN alloying}

\author{Sukriti Manna$^1$, Kevin R. Talley$^{2,3}$, Prashun Gorai$^{2,3}$, John Mangum$^2$, Andriy Zakutayev$^3$, Geoff L. Brennecka$^2$, Vladan Stevanovi\'{c}$^{2,3}$, and Cristian V. Ciobanu$^1$\footnote{Corresponding author, email: cciobanu@mines.edu}}
\affiliation{$^1$Dept. of Mechanical Engineering, Colorado School of Mines, Golden, Colorado 80401, USA\\
$^2$Dept. of Metallurgical and Materials Engineering, Colorado School of Mines, Golden, Colorado 80401, USA\\
$^3$National Renewable Energy Laboratory, Golden, CO 80401, USA}
\date{\today}% It is always \today, today,

\begin{abstract}

Since AlN has emerged as an important piezoelectric material for a wide variety of applications, efforts have been made to increase its piezoelectric response via alloying with transition metals that can substitute for Al in the wurtzite lattice. Herein, we report density functional theory calculations of structure and properties of the Cr-AlN system for Cr concentrations ranging from zero to beyond the wurtzite-rocksalt transition point. By studying the different contributions to the longitudinal piezoelectric coefficient, we propose that the physical origin of the enhanced piezoelectricity in Cr$_x$Al$_{1-x}$N alloys is the increase of the internal parameter $u$ of the wurtzite structure upon substitution of Al with the larger Cr ions.
Among a set of wurtzite-structured materials, we have found that
Cr-AlN has the most sensitive piezoelectric coefficient with respect to alloying concentration.
Based on these results, we propose that Cr-AlN is a viable piezoelectric material whose properties can be tuned via Cr composition. We support this proposal by combinatorial synthesis experiments, which show that Cr can be incorporated in the AlN lattice up to 30\% before a detectable transition to rocksalt occurs.
At this Cr content, the piezoelectric modulus $d_{33}$ is approximately four times larger than that of pure AlN.
This finding, combined with the relative ease of synthesis under non-equilibrium conditions, may propel Cr-AlN as a prime piezoelectric material for
applications such as resonators and acoustic wave generators.

\end{abstract}

%\keywords{piezoelectricity, mixing enthalpy, cystallinity, thin film}
\maketitle

%\tableofcontents

\section{Introduction}

Aluminum nitride has emerged as an important material
for micro-electromechanical (MEMS) based systems\cite{fu2017advances,muralt2008recent} such as
surface and bulk acoustic resonators,\cite{fu2017advances,loebl2003piezoelectric} atomic force microscopy (AFM) cantilevers,\cite{fu2017advances} accelerometers,\cite{gerfers2007sub,wang2017mems} oscillators,\cite{zuo20101} resonators for energy harvesting,\cite{wang2017aln,elfrink2009vibration}
and band-pass filters.\cite{yang2003highly} The advantages of using AlN in MEMS devices include
metal$-$oxide$-$semiconductor (CMOS) compatibility, high thermal conductivity, and high temperature stability. In addition, its low permittivity
and high mechanical stiffness are particularly important for resonantor applications.\cite{fu2017advances,muralt2017aln}
However, the piezoelectric constants of AlN thin films are lower than those of other commonly used piezoelectric materials.
For example, the out-of-plane piezoelectric strain modulus\cite{nomenclature} $d_{33}$ of reactively sputtered AlN films is reported to be 5.5 pC/N,  whereas  $d_{33}$ for ZnO
can be at least twice as large,\cite{kang2017enhanced} and PZT films can be over 100 pC/N.\cite{muralt2008recent}

It is therefore desirable to find ways to increase the piezoelectric response of AlN in
order to integrate AlN-based devices into existing and new systems. A common way to engineer piezoelectric properties of AlN is by alloying with transition metal nitrides (Sc, Y, others), which can lead to a several-fold increase in the field-induced strain via  increases in the longitudinal piezoelectric coefficient $e_{33}$ and simultaneous
decreases in the longitudinal elastic stiffness $C_{33}$.\cite{akiyama2009enhancement,caro2015piezoelectric,manna2017tuning,tasnadi2010origin}
In the case of ScN alloying, the origins of this response have been studied,\cite{tasnadi2010origin} and it is presumed that
other such systems which also involve AlN alloyed with rocksalt-structured end members are similar:
as the content of the rocksalt end member in the alloy increases, the accompanying structural frustration enables
a greater piezoelectric response.
This structural frustration, however, is also accompanied by
thermodynamic driving forces for phase separation\cite{hoglund2010wurtzite} which, with increased alloy concentration, lead to the
destruction of the piezoelectric response upon transition to
the (centrosymmetric, cubic) rocksalt structure.
The experimental realization of large alloy contents without phase separation or severe degradation of film texture and crystalline quality can be quite difficult,\cite{hoglund2010wurtzite, mayrhofer2015microstructure} even when
using non-equilibrium deposition processes such as sputtering. Thus, it is desirable to find alloy systems for which
the structural transition from wurtzite to rocksalt occurs at low alloying concentrations since these may be more easily synthesized and more stable, while also (hypothetically) providing comparable property enhancements as those observed in the more-studied Sc-AlN alloy system.
Among the AlN-based systems presently accessible experimentally, Cr-AlN has the lowest transition composition between the wurtzite and rocksalt structures,
occurring at approximately 25\% CrN concentration.\cite{mayrhofer2008structure, holec2010pressure} This motivates the investigation of the
piezoelectric properties of the Cr$_x$Al$_{1-x}$N system, which we also refer to, for simplicity, in terms of Cr substitution for Al.

In this article, we study Cr-substituted AlN using density functional theory (DFT) calculations of
structural, mechanical, and piezoelectric properties. Given that Cr has unpaired $d$ electrons, a challenge to overcome in these calculations
is the simulation of a truly representative random distribution of the spins of Cr ions, whose placement in the AlN lattice involves not
only chemical disordering, but spin disordering as well.
Among a set of wurtzite-based materials, we have found that
Cr-doped AlN is the alloy whose piezoelectric stress coefficient $e_{33}$ is the most sensitive to alloying concentration
and also has the lowest wurtzite-to-rocksalt transition composition.
The key factor leading to the enhanced piezoelectricity in Cr$_x$Al$_{1-x}$N alloys is the ionic contribution to the coefficient $e_{33}$; this ionic contribution is increased through the internal $u$ parameter of the wurtzite structure when alloyed with the (larger) Cr ions.
Therefore, we propose Cr$_x$Al$_{1-x}$N as a viable piezoelectric material with properties that can be tuned via Cr composition.
To further support this proposal, we have performed combinatorial synthesis and subsequent characterization of Cr$_x$Al$_{1-x}$N films,
and have showed that Cr can be incorporated in the AlN lattice up to 30\% before a detectable transition to rocksalt occurs.
At this Cr content, the piezoelectric modulus $d_{33}$ is four times larger than that of AlN. Pending future device fabrication and accurate
measurements of properties and device performance, this significant increase in $d_{33}$ can propel Cr-AlN to be the
choice material for applications such as resonators, GHz telecommunications, or acoustic wave generators.

\section{Methods}

\subsection{Paramagnetic Representation of Cr-AlN Alloys}

\begin{figure}[h]
	\includegraphics[width=7cm,angle=270]{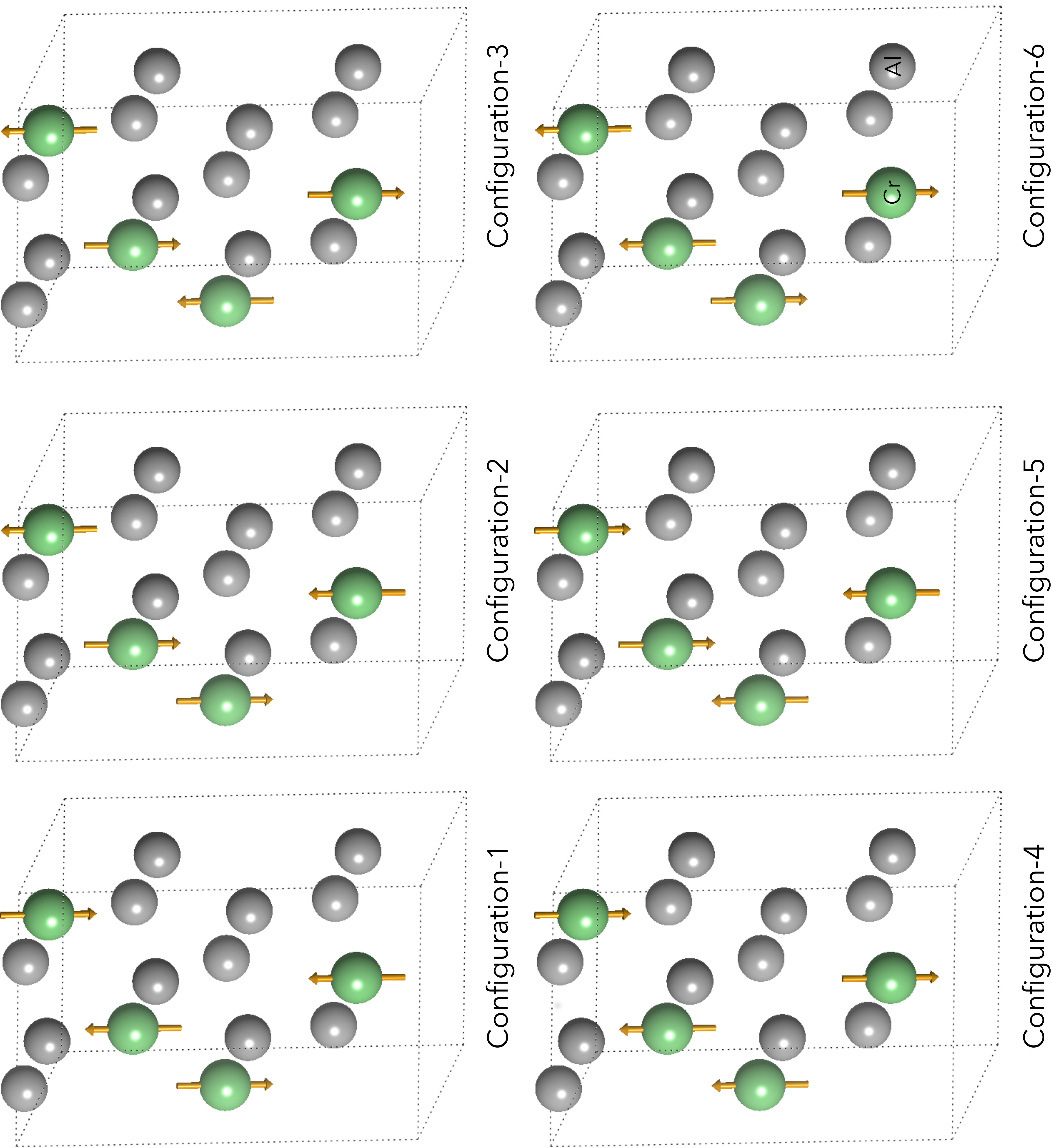}
	\caption{Schematics of cation sublattice of Cr$_{x}$Al$_{1-x}$N alloy. Al (Cr) sites are shown as gray (green) spheres. At a given Cr concentration, the Cr sites of each configuration have a different and random spin initialization with zero total spin in order to capture the paramagnetic state.}
	\label{schematics-spin}
\end{figure}

Starting with a computational supercell of wurtzite AlN, any desired Cr concentration is realized by substituting a corresponding number of Al ions with Cr ions in the cation sub-lattice. In order to realistically simulate the chemical disorder of actual Cr-AlN alloys while maintaining a tractable size for the computational cell, we use special quasirandom structures (SQS).\cite{zunger1990special,van2009multicomponent,van2013efficient}
The Cr$^{3+}$ ions have unpaired $d$ electrons, which require spin-polarized DFT calculations. Another important aspect of the
calculations is that the Cr$_x$Al$_{1-x}$N alloys are paramagnetic,\cite{mayrhofer2008structure,endo2007crystal,endo2005magnetic} and this state has to be captured explicitly in the DFT calculations.  Therefore, in addition to the configurational disorder simulated via SQS, the paramagnetic state requires truly random configurations for the spins associated with the  Cr$^{3+}$ ions.\cite{alling2010effect,abrikosov2016recent}
However, as shown by Abrikosov {\em et al.},\cite{abrikosov2016recent} the paramagnetic state can be approximated by using
disordered, collinear, static spins because such state yields zero spin-spin correlation functions.
To represent the paramagnetic state of Cr$_x$Al$_{1-x}$N, for a given alloy structure with $n$ Cr sites, we performed a minimum of $n \choose 2$ and maximum 20 calculations. In these calculations, the spins on Cr sites are randomly initialized subject to the restriction of zero total spin for each concentration and each SQS structure. An example of such a random distribution of initial spins is illustrated in Figure~\ref{schematics-spin} for $x$ = 25\% Cr concentration.

\subsection{Details of the DFT Calculations}

Structural optimizations and calculations of piezoelectric and elastic constants were carried out using the Vienna Ab-initio Simulation Package (VASP),\cite{kresse1996efficiency} with projector augmented waves (PAW) in the generalized gradient approximation using the Perdew-Burke-Ernzerhof (PBE) exchange-correlation function\cite{perdew1996generalized} and an on-site Hubbard term\cite{dudarev1998electron} $U$ for the Cr $3d$ states. The plane wave cutoff energy was set to 540 eV in all calculations.
For the wurtzite structures, we have used $4\times4\times 2$ (128 atoms) and $2\times2\times2$ (32 atoms) SQS supercells; for the rocksalt structures, the computations were carried out on $2\times 2 \times 2$ (64 atoms) SQS supercells. Brillouin zone sampling was performed by employing $1\times 1\times 1$ and $2 \times2\times 2$ Monkhorst-Pack\cite{monkhorst1976special} $k$-point meshes for the wurzite and rocksalt structures, respectively, with the origin set at the $\Gamma$ point in each case. Piezoelectric coefficients were calculated using density functional perturbation theory, and the elastic constants were computed by finite differences.\cite{gonze1997dynamical,wu2005systematic} The on-site Coulomb interaction for Cr atoms was set at 3 eV, through a Dudarev approach.\cite{dudarev1998electron} Before performing the calculations for elastic and piezoelectric constants, we performed cell shape, volume, and ionic relaxations in order to obtain the equilibrium lattice parameters and ionic positions at each particular Cr concentration and SQS alloy.

\subsection{Experimental Procedures}
Combinatorial synthesis of Cr$_x$Al$_{1-x}$N films was performed through reactive physical vapor deposition (PVD). Two inch diameter circular aluminum (99.9999\%) and chromium (99.999\%) metallic targets were arranged at 45$^{\rm o}$ angles measured from the normal to a plasma-cleaned Si(100) substrate inside a custom vacuum system with a base pressure of $5\times 10^{-6}$ torr. Magnetron RF sputtering with a power of 60 W for aluminum targets and  40 W for the chromium targets was performed at a deposition pressure of $3\times 10^{-3}$ torr, with 8 sccm of argon and 4 sccm of nitrogen, and a substrate temperature of 400 $^{\rm o}$C. Aluminum glow discharges were oriented opposite to each other,
with the chromium target perpendicular to both, resulting in a film library with a compositional range in one direction.\cite{config1,config2}
Each sample library was subdivided into eleven regions across the composition gradient, which were subsequently characterized by x-ray diffraction (XRD) and x-ray fluorescence (XRF), performed on a Bruker D8 Discovery diffractometer with a 2D area detector in a theta-2theta configuration and a Fischer XUV vacuum x-ray spectrometer, respectively.

\section{Results and Discussion}

\subsection{Enthalpy of Mixing}
\begin{figure}[!htbp]
\begin{center}
\includegraphics[width=7cm]{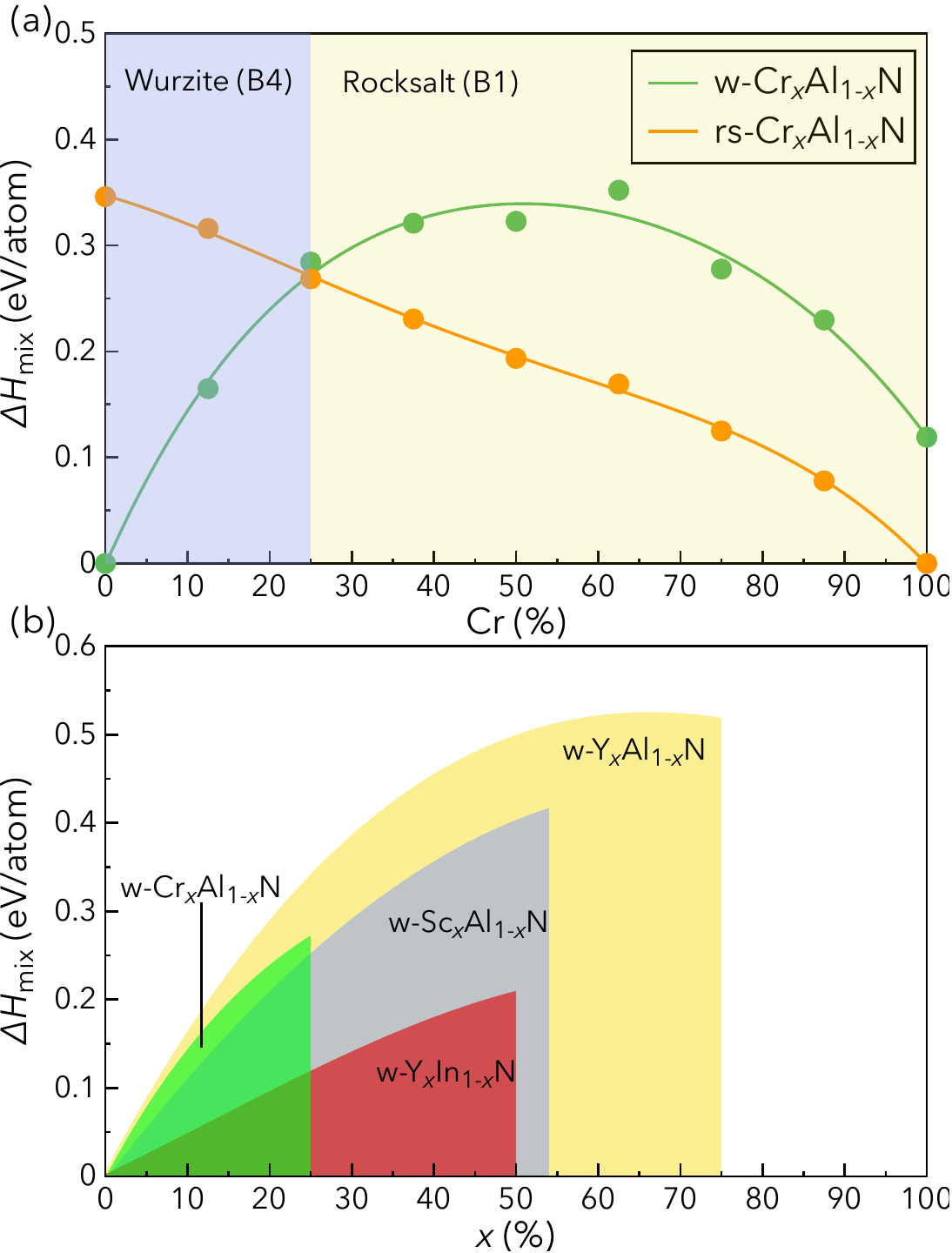}
\caption{(a) DFT-calculated mixing enthalpies of the wurtzite and rocksalt phases of Cr$_{x}$Al$_{1-x}$N as functions of Cr concentration.  
 (b)  Calculated  mixing enthalpies for several wurzite-based nitride alloys grown experimentally.}
\label{enthalpy}
\end{center}
\end{figure}

The enthalpy of mixing as a function of the Cr concentration $x$, at zero pressure, is defined with respect to the pure wurzite-AlN and rocksalt-CrN phases via
\begin{equation}
\Delta H_{\text{mix}}(x) = E_{\text{Cr}_{x}\text{Al}_{1-x}\text{N}} - xE_{\text{rs}\text{-CrN}} -(1-x)E_{\text{w}\text{-AlN}},
\end{equation}
where $E_{\text{Cr}_{x}\text{Al}_{1-x}\text{N}}$, $E_{\text{rs}-\text{CrN}}$, and $E_{\text{w}-\text{AlN}}$ are the total energies per atom of the SQS alloy, pure AlN phase, and pure CrN phase, respectively . The DFT calculated mixing enthalpies for the wurtzite and rocksalt phases of Cr$_{x}$Al$_{1-x}$N are shown in Figure~\ref{enthalpy}(a). The wurzite phase is found to be favorable up to $x=0.25$, beyond which rocksalt alloys are stable; this wurzite to rocksalt phase transition point is consistent with previous experimental observations and other theoretical predictions.\cite{mayrhofer2008structure, holec2010pressure}

We have compared the mixing enthalpy of the Cr$_x$Al$_{1-x}$N alloys with that of several other common
wurzite-based nitrides,\cite{akiyama2009enhancement,tholander2016ab,vzukauskaite2012yxal1,manna2017tuning,mayrhofer2015microstructure} Sc$_{x}$Al$_{1-x}$N, Y$_{x}$Al$_{1-x}$N, and Y$_{x}$In$_{1-x}$N, with the results shown in Figure~\ref{enthalpy}(b).
The mixing enthalpies are positive for all cases, meaning that the alloying of AlN or InN with their respective end members is an endothermic process. In practice, these alloys are formed as disordered solid solutions obtained
using physical vapor deposition techniques operating at relatively low substrate temperatures  because of the energetic plasmas involved.\cite{hoglund2010wurtzite,luo2009influence}
Figure~\ref{enthalpy}(b) shows that the mixing enthalpy in Cr$_x$Al$_{1-x}$N lies between values corresponding to other systems synthesized experimentally, hence Cr$_x$Al$_{1-x}$N is no more difficult to synthesize than the others. More importantly, the enthalpy calculations show that the transition to rocksalt occurs at the lowest alloy concentration across the wurtzite systems considered, which is important for achieving maximum piezoresponse-enhancing structural frustration with a minimum of dopant concentration in order to retain the single-phase wurtzite.

\begin{figure*}[!htbp]
\begin{center}
\includegraphics[width=13cm,angle=270]{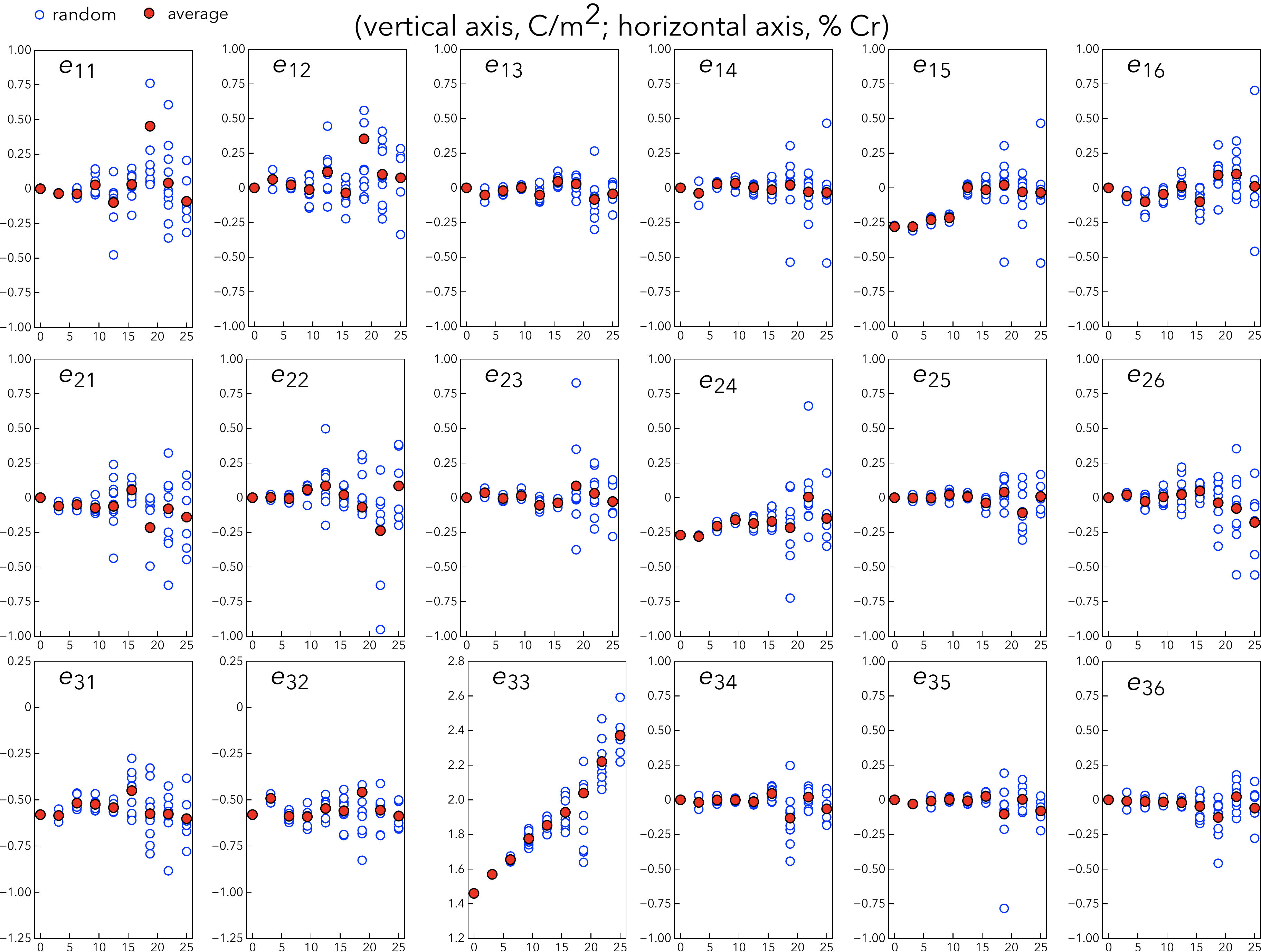}
\caption{The 18 components of the piezoelectric tensor calculated for SQS supercells with different spin configurations and compositions of Cr. At each Cr composition, the open circles represent data for each SQS cell used, while the red solid circles represent the average values across the random initial configurations.}
\label{all-e}
\end{center}
\end{figure*}

\begin{figure*}[!htbp]
\includegraphics[width=10cm,angle=270]{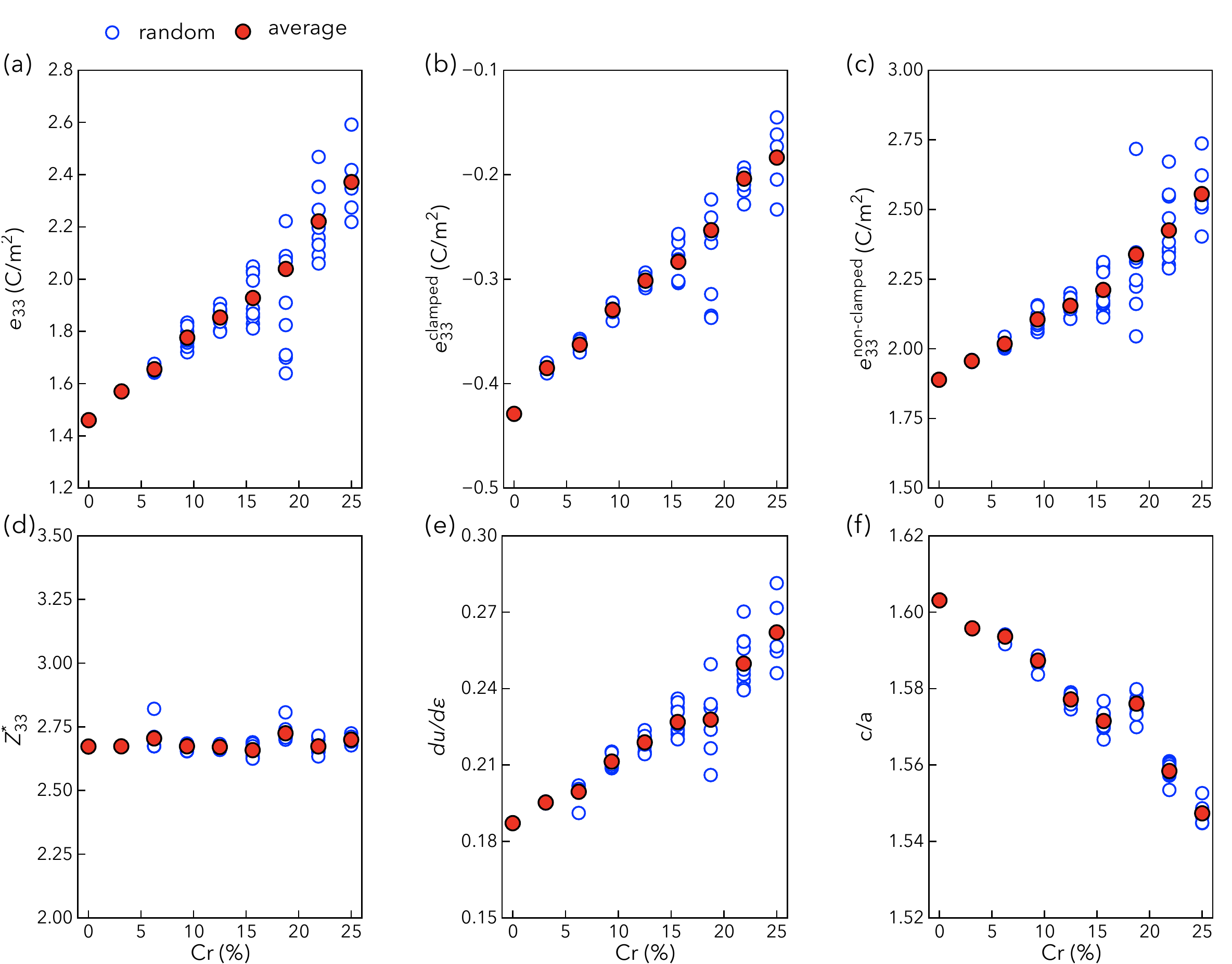}
\caption{Variation of (a) $e_{33}^{total}$, (b) $e_{33}^{clamped}$ component of $e_{33}$, (c)  $e_{33}^{non clamped}$, (d) 33 component of the Born effective charge tensor, (e) strain sensitivity of internal parameter, (f) $c/a$ ratio with respect to Cr addition.}
\label{e33-details}
\end{figure*}

\begin{figure}[h]
	\includegraphics[width=5cm,angle=270]{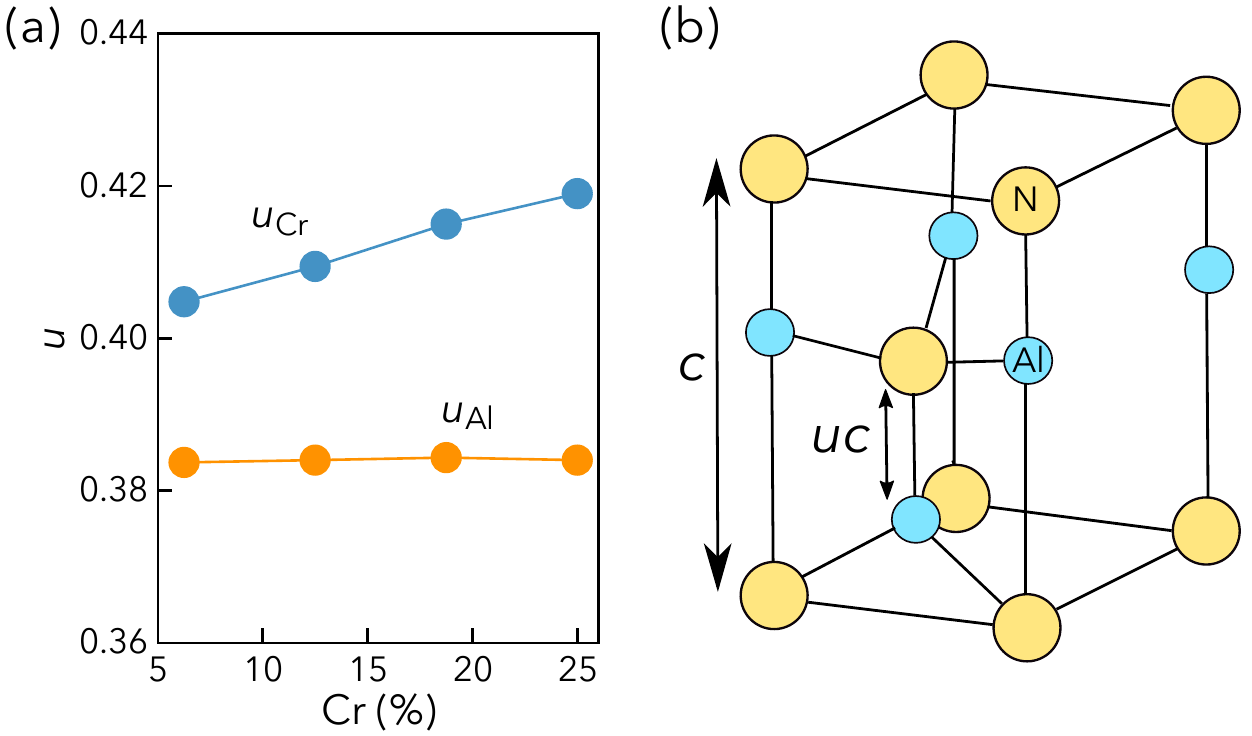}
	\caption{(a) Variation of internal parameter, u with Cr addition. (b) Crystal structure of wurzite AlN.}
	\label{tetrahedra}
\end{figure}

\subsection{Piezoelectric Stress Coefficients}
The piezoelectric coefficients $e_{ij}$ for different spin configurations in SQS supercells with the same Cr content are shown in Figure~\ref{all-e}.
For clarity, the panels in Figure~\ref{all-e} are arranged in the same fashion as the piezoelectric tensor when represented as a matrix in Voigt notation. The vertical scale is the same for all coefficients except $e_{33}, e_{31}$, and $e_{32}$. The scatter in the results corresponds to different SQS supercells at each  Cr concentration; this is an effect of the finite size of the system, in which local distortions around Cr atoms lead to small variations of the lattice constants and angles.
It is for this reason that we average the SQS results at each Cr concentration, thereby obtaining smoother variations of the piezoelectric coefficients. At  25\% Cr, the value of $e_{33}$ becomes $\sim$1.7 times larger than that corresponding to pure AlN.

The piezoelectric coefficient $e_{33}$ of wurzite Cr$_x$Al$_{1-x}$N is shown in Figure~\ref{e33-details}(a) as a function of Cr concentration,
and can be written as\citep{bernardini1997spontaneous}
\begin{equation}
e_{33}(x) = e_{33}^{\text{clamped}}(x) + e_{33}^{\text{non-clamped}}(x),
\label{eqe33partition}
\end{equation}
in which $e_{33}^{\text{clamped}}(x)$  describes the electronic response to strain and is evaluated by freezing the internal atomic coordinates at their equilibrium
positions. The term $e_{33}^{\text{non-clamped}}(x)$ is due to changes in internal coordinates, and is given by
\begin{equation}
e_{33}^{\text{non-clamped}}(x) = \frac{4eZ_{33}^{*}(x)}{\sqrt{3} a(x)^{2}}\frac{du(x)}{d\epsilon}
\label{nclamped}
\end{equation}
where $e$ is the (positive) electron charge, $a(x)$ is the equilibrium lattice constant, $u(x)$ is the internal parameter of the wurtzite, $Z_{33}^{*}(x)$ is the dynamical Born charge in units of $e$, and $\epsilon$ is the macroscopic applied strain. $e_{33}^{\text{non-clamped}}(x)$ describes the piezoelectric response coming from the displacements of internal atomic coordinates produced by the macroscopic strain.
Based on Eqs.~(\ref{eqe33partition}) and (\ref{nclamped}), panels (b) through (f) in Figure~\ref{e33-details} show the different relevant quantities contributing to $e_{33}$ in order to identify the main factors responsible for the increase of piezoelectric response with Cr addition.
Direct inspection of Figures~\ref{e33-details}(a-c) indicates that the main contribution to the increase of $e_{33}$ comes from the non-clamped ionic part, Figure~\ref{e33-details}(c). Since the Born charge $Z_{33}^{*}$ [Figure~\ref{e33-details}(d)] is practically constant, the key factor that leads to increasing the piezoelectric coefficient is the strain sensitivity $du/d\epsilon$ of the internal parameter $u$ [Figure~\ref{e33-details}(e)].

Although the internal parameter $u$ is an average value across the entire supercell, the individual average $u$ parameters can also be determined  separately for AlN and CrN tetrahedra [Figure~\ref{tetrahedra}(a,b)]. The internal parameter $u$ of AlN tetrahedra [Figure~\ref{tetrahedra}(b)] does not change significantly, while that of the CrN tetrahedra grows approximately linearly with Cr concentration [Figure~\ref{tetrahedra}(a)].
In an alloy system where AlN tetrahedra are the majority, this variation can be understood based on (i) the fact that the
ionic radius of Cr is about 10\% larger than that of Al, and (b) the increase in Cr concentration will lead to average
 $u$ parameters mimicking the variation of the $u$ parameter corresponding to CrN tetrahedra.

\subsection{Comparison with Other Wurtzite-Based Alloys}
\begin{figure}[h]
\includegraphics[width=7cm]{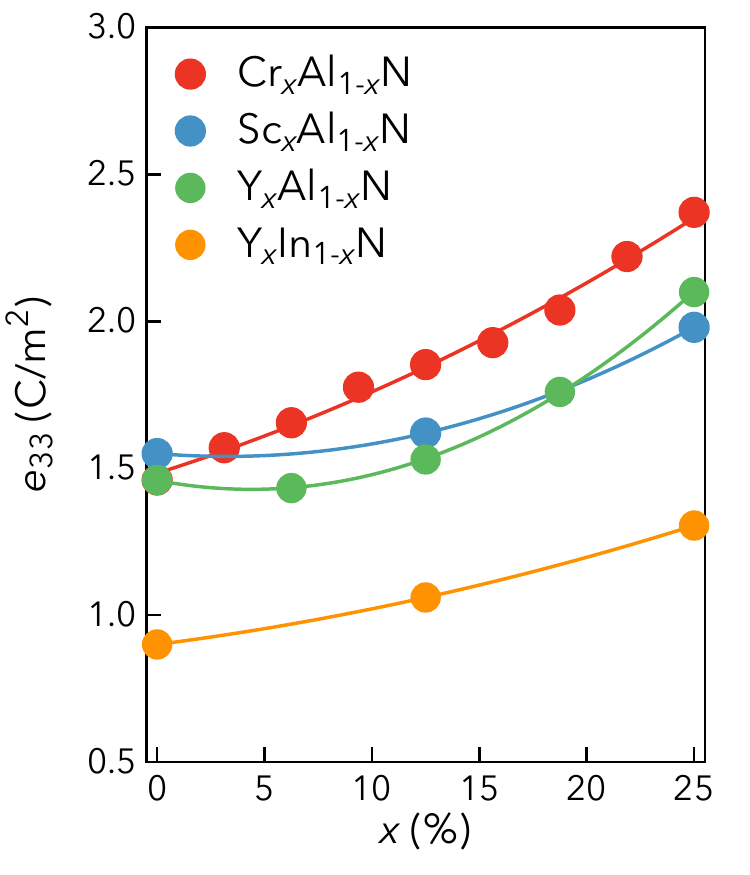}
\caption{Comparison of change in $e_{33}$ with addition of different transition metals for $x\leq$ 25\% regime.}
\label{e33-compare}
\end{figure}

The results from calculations of the piezoelectric properties of Cr$_x$Al$_{1-x}$N with $x$ from 0 to 25\% Cr are plotted in Figure~\ref{e33-compare},
together with the calculated values for Sc$_x$Al$_{1-x}$N, Y$_x$Al$_{1-x}$N, and Y$_x$In$_{1-x}$N.
In Cr$_x$Al$_{1-x}$N,  $e_{33}$ increases rapidly from 1.46 to 2.40 C/m$^{2}$ for Cr concentrations from 0 to 25\%.
For all other alloys considered, the increase is smaller in the same interval of solute concentration:
for Sc$_x$Al$_{1-x}$N, Y$_x$Al$_{1-x}$N, and Y$_x$In$_{1-x}$N, $e_{33}$ increases, respectively, from 1.55 to 1.9 C/m$^{2}$, 1.55 to 1.7 C/m$^{2}$,
and 0.9 to 1.2 C/m$^{2}$.
Within the $x\leq$ 25\% range, Cr is more effective than any of the other studied transition elements in improving piezoelectric response of AlN-based alloys.

\begin{figure}[h]
\includegraphics[width=6cm]{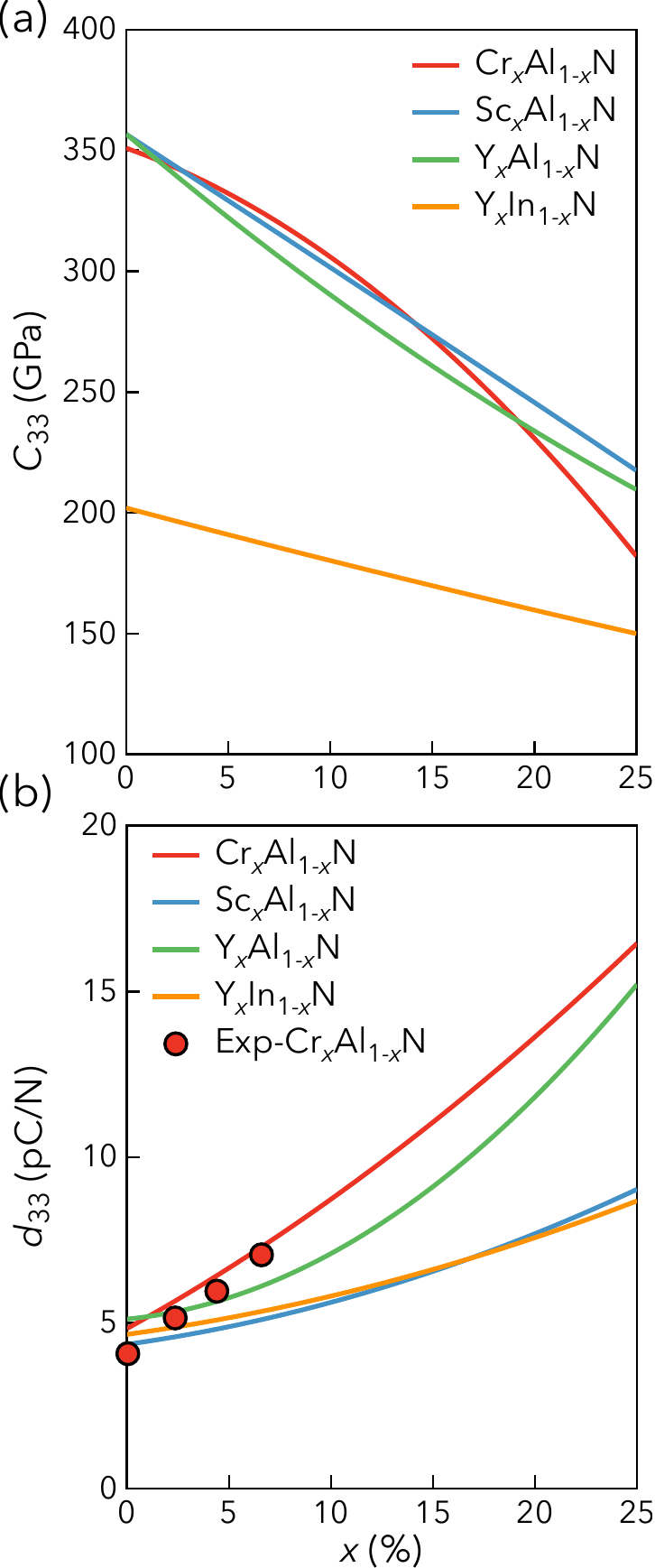}
\caption{Variation of (a) $C_{33}$ and (b) $d_{33}$ for several nitride-based wurtzite alloys.}
\label{C33d33}
\end{figure}

\begin{figure}
\includegraphics[width=7.5cm]{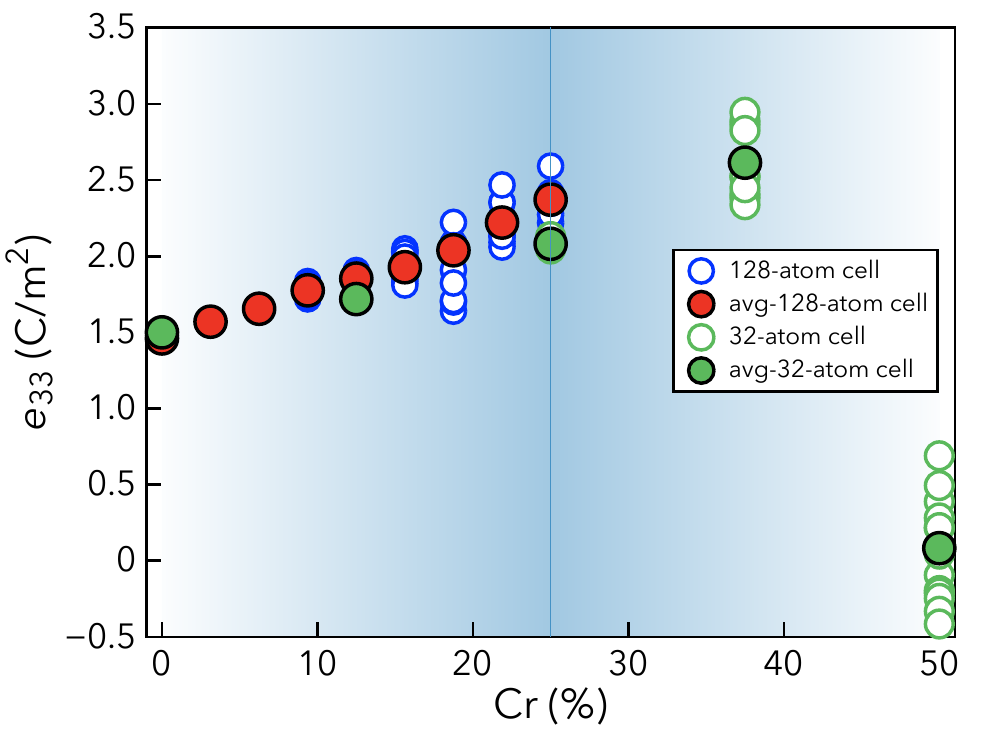}
\caption{Variation of $e_{33}$ as function of Cr concentration for wurtzite phase alloys up to 50\%.}
\label{e33-25}
\end{figure}

The experimentally measurable  property is $d_{33}$, which is commonly known as piezoelectric strain modulus
and relates the electric polarization vector with stress.
The relationship between the piezoelectric strain and stress moduli is\cite{nye1985physical}
\begin{equation}
d_{ij} = \sum_{k=1}^{6} e_{ik}(C^{-1})_{kj},
\label{eC}
\end{equation}
where $C_{ij}$ are the elements of the stiffness tensor in Voigt notation.
The variation of the elastic constant $C_{33}$ in Cr$_x$Al$_{1-x}$N with $x$ is shown in Figure~\ref{C33d33}(a),
along with the other systems considered here.
For all of these wurtzite-based piezoelectrics, the increase in piezoelectric response with alloying element concentration is accompanied by mechanical softening (decrease in $C_{33}$).
From Eq.~(\ref{eC}), it follows that the
increase in $e_{33}$ (Figure~\ref{e33-compare}) and the mechanical softening [Figure~\ref{C33d33}(a)]
cooperate to lead to the increase of $d_{33}$ values with alloy concentration $x$.
Our calculated $d_{33}$ values for Cr$_x$Al$_{1-x}$N are in good agreement with experimental data from Ref.~\onlinecite{luo2009influence} [Figure~\ref{C33d33}(b)]
for Cr concentrations up to 6.3\%. Beyond this concentration, Luo et al.\cite{luo2009influence} report a drop in the $d_{33}$ values of their films, which is attributed 
to changes in film texture.
We have also extended our calculations of piezoelectric coefficients beyond 25\% Cr composition in wurtzite structures.
Figure~\ref{e33-25} shows that $e_{33}$ continues to increase at least up to 37.5\% Cr.
The calculations done at 50\% Cr, which start with wurtzite SQS configurations, evolve into rocksalt configurations during relaxation, which explains
the decrease of $e_{33}$ to zero in Figure~\ref{e33-25}.

\begin{table}[htbp]
	\centering
	\caption{Piezoelectric properties of AlN and a few wurtzite alloys for piezoelectric device applications.}
	\label{table:properties}
	\begin{tabular}{ l  c  c  c  c }
		\hline \hline
		Material  & $e_{33}$ (C/m$^2)$   &  $d_{33}$ (pC/N) &  Refs. \\
		\hline
        AlN             & 1.55  & 4.5-5.3 &   \onlinecite{fu2017advances,muralt2008recent,muralt2017aln}  \\
        Sc-AlN, 10\% Sc & 1.61  & 7.8     &   \onlinecite{appl8,appl9}\\
        Y-AlN, 6\% Y    & 1.5   & 4.0     &  \onlinecite{mayrhofer2015microstructure} \\
        Y-InN, 14\% Y   & 1.1   & 5.1     &  \onlinecite{tholander2016ab} \\
       \hline	
          {\em this work:}           &       &      &     \\
        Cr-AlN, 12.5\% Cr            & 1.84  & 9.86 &     \\
        Cr-AlN, 25.0\% Cr           & 2.35  & 16.45 &     \\
        Cr-AlN, 30.0\% Cr           & 2.59  & 19.52 &     \\

		\hline \hline

	\end{tabular}

\end{table}

It is worthwhile to compare the performance of several AlN wurzite-based materials
for their use in applications. These applications, which are mainly resonators, ultrasound wave generators, GHz telecommunications, FBAR devices, bulk or surface acoustic generators, and biosensors, lead to a multitude of application-specific figures of merit for different utilization modes of the piezoelectric material.
However, most figures of merit rely on the piezoelectric properties $e_{33}$ and $d_{33}$, both of which in general should be as large as possible for increased piezoelectric
device sensitivity.
The most used wurtzite material for these applications is AlN, although there are several other options as well (refer to Table I).
Alloying with ScN is promising in that it offers an increased $d_{33}$ for about 10\% Sc concentration; larger
Sc concentrations are possible, but the growth process becomes more costly and the material is likely to
lose texture with increased Sc content. Options such as alloying with YN offer marginal improvement at 6\% Y content,
and YN-doped InN (14\% Y content) fares similarly (Table I).
Our results indicate that CrN alloying of AlN can reach superior values for the piezoelectric
properties, nearly quadrupling the value of $d_{33}$ (Table I) with respect to AlN.
The fact that the transition point is the lowest (Figure~\ref{enthalpy}) of all wurtzite-based materials relevant for
the technologies mentioned above, makes the CrN alloying easier compared with the other materials (which require higher alloy content)
 and hence renders Cr-AlN  a prime candidate for synthesis of new, CrN-alloyed piezoelectrics for resonators and acoustic generators.
As we shall see in Sec. III.D, the non-equilibrium growth techniques can bring Cr content past the transition  point
without significant formation of the (non-piezoelectric) rocksalt phase. Consequently,
 the piezoelectric properties are expected to be significantly better than those of AlN, especially $d_{33}$ (refer to Table~\ref{table:properties}). Indeed,
this is born out in experiments (data points in Figure~\ref{C33d33}b).
Measurements
 of figures of merit for specific device configurations will be needed in the future,
 as those require not only combinations of elastic and piezoelectric properties, but dielectric properties as well.\cite{manna2017tuning}

\subsection{Experimental Results}
\begin{figure}[h]
\begin{center}
\includegraphics[width=8.5cm]{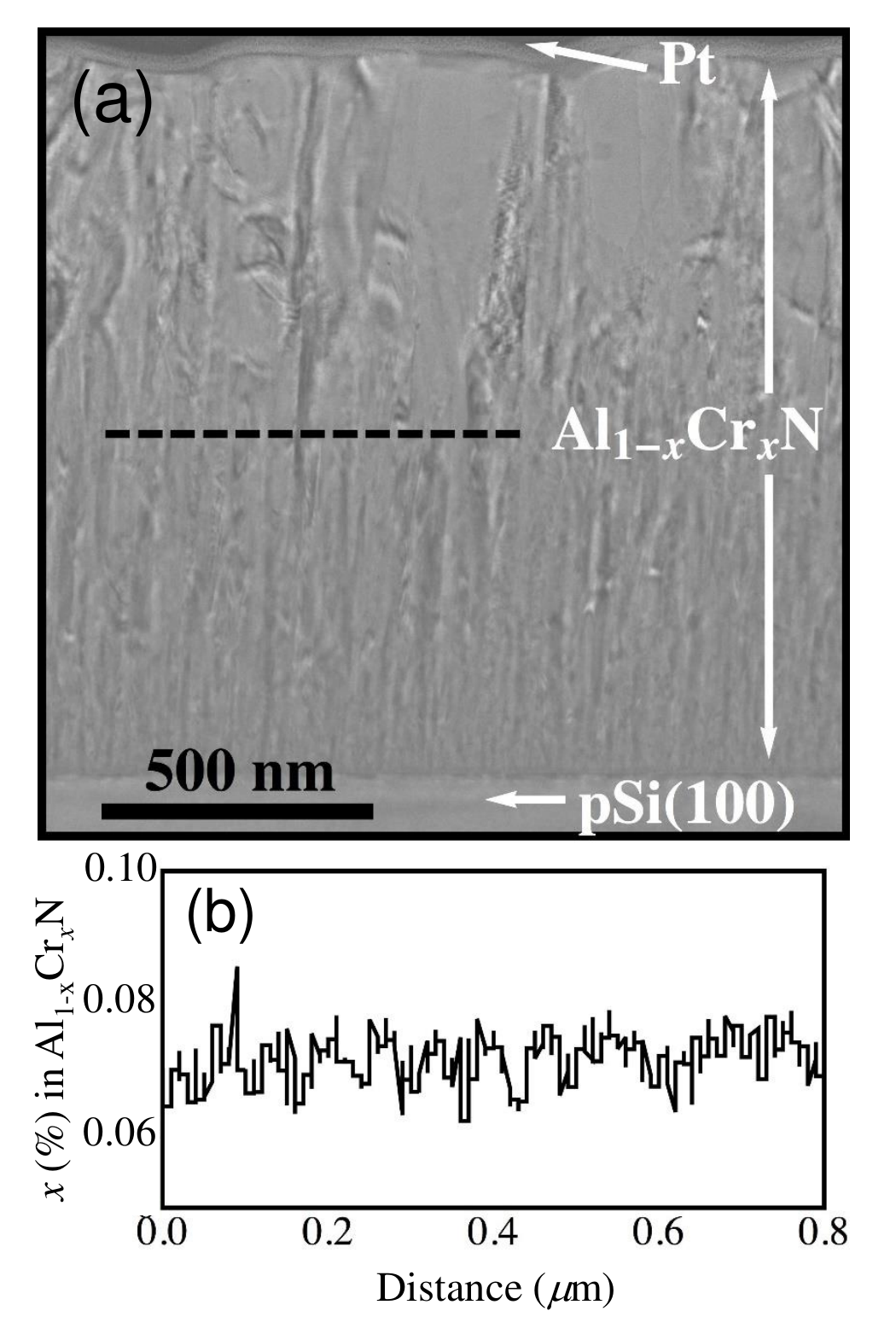}
\caption{ Representative transmission electron microscopy (TEM) image (a) and an energy dispersive spectroscopy (EDS) line scan (b) of an (Al$_{1-x}$Cr$_x$)N film cross section containing $\sim$7\% Cr, confirming the incorporation of Cr into wurtzite solid solution. EDS data were collected along the dashed black line shown in panel (a).}
\label{TEM}
\end{center}
\end{figure}

\begin{figure}[h]
	\includegraphics[width=8.5cm]{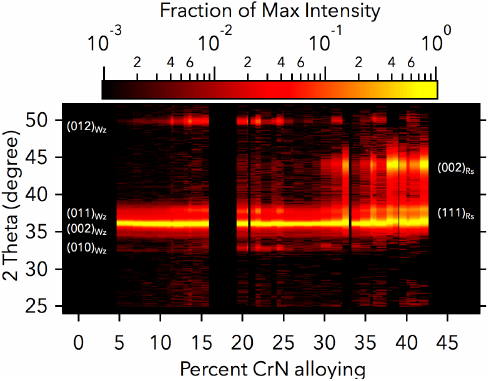}
	\caption{X-ray diffraction patterns of the thin film combinatorial libraries plotted against the film composition, with comparison to the patterns for wurtzite (WZ) \cite{eddine1977etude} and rocksalt(RS)\cite{wyckoff1960crystal} structures. For alloying content $x<30\%$, the films grow predominantly with the
wurtzite structure. At higher Cr concentrations, $x>30\%$, both rocksalt and wurtzite phases are detected, and the wurtzite exhibits degraded texture. No films were produced with compositions in regions where no intensity is shown. }
	\label{kevin}
\end{figure}
To bring experimental support to our proposal that the Cr-AlN system can become a key piezoelectric material to replace AlN and perhaps even ZnO for future applications, we
have to ensure that the texture obtained during growth is stable for sufficiently high CrN concentrations.
After synthesizing Cr-AlN alloys through reactive PVD, we have performed transmission electron microscopy (TEM) analysis of
the films grown in order to check for textural integrity (i.e., grains oriented primarily with the $c$ axis close to the surface normal)
and for the onset of the rocksalt phase.
At CrN content below 25\% [the theoretical boundary shown in Figure \ref{enthalpy}(a)], our films display  no significant texture variations.
For example, Figure 9(a) shows a typical TEM micrograph wherein texture is preserved over the film thickness.
Additionally,  our energy dispersive spectrocopy (EDS) characterization shows nearly constant Cr content
through the sample [Figure \ref{TEM}(b)].
Further characterization by XRD was performed for all CrN compositions in the combinatorially synthesized films.
Figure~\ref{kevin} shows the XRD results for the 88 discrete Cr$_x$Al$_{1-x}$N compositions produced in an effort to test the possibilities
of synthesizing alloys in a wide range of concentrations, including alloys beyond the wurtzite-to-rocksalt transition point.
At low alloying levels, the films grow exclusively with the wurtzite structure and a $(002)$ preferred orientation, as indicated by the dominant presence of the wurtzite $(002)$ diffraction peak (Figure~\ref{kevin}, left side).
Films grown by reaction PVD under the conditions used here accept chromium into the wurtzite lattice and grow primarily with the ideal $(002)$ orientation.
With increased CrN content, the wurtzite $(012)$ and $(010)$  peaks appear, indicating some deviations from the original, and still predominant $(002)$ orientation of the film.
The metastability of this alloy is overcome at an approximate composition of $x \simeq  30$\%, where the polycrystalline rocksalt phase appears, as revealed by the rocksalt $(002)$ and $(111)$ peaks (Figure~\ref{kevin}, right hand side).
These experimental results show that wurtzite Cr$_x$Al$_{1-x}$N solid solutions can be synthesized without observable phase separation up to concentrations of 30\% Cr. Wurtzite material still exists at global compositions beyond 30\%, but in a wurtzite-rocksalt phase mixture, which will diminish the piezoelectric properties because of the presence of a significant amount of centrosymmetric rocksalt phase in the mixture.

There are few studies of Cr alloyed into wurtzite AlN,\cite{luo2009influence, felmetsger2011, endo2005magnetic} reporting Cr-doped
alloys grown by magnetron sputtering. The Cr concentration previously attained is below 10\%, although
the limits of Cr alloying were not actually tested in the previous reports.\cite{luo2009influence, felmetsger2011, endo2005magnetic} Our combinatorial synthesis results show that Cr can be doped into the wurtzite lattice up to 11\% before the predominant (002) film texture starts to change, and up to 30\% before the rocksalt phase appears.

\section{Concluding Remarks}

By using a physical representation of the paramagnetic state of substitutional Cr
 in a wurtzite AlN matrix and performing the necessary averaging
 over spin configurations at each Cr concentration, we computed the structural, mechanical, and piezoelectric properties of Cr-AlN alloys.
Our combinatorial synthesis experiments showed that Cr-AlN are relatively easy to synthesize,
and also showed that the reactive PVD procedure resulted in  Cr-AlN alloys retaining  the wurtzite structure for alloying concentrations up to 30\% Cr.
Remarkably, our DFT calculations of piezoelectric properties revealed that for 12.5\% Cr $d_{33}$ is twice that of pure AlN, and for 
30\% Cr this modulus is about four times larger than that of AlN.

From a technological standpoint, this finding should make Cr-AlN the prime candidate to
replace the current-wurtzite based materials in resonators and acoustic wave generators.
The larger piezoelectric response (than AlN) may lead to smaller power consumption
and perhaps even to avenues to further miniaturize various devices.
While the substitutional alloying with Cr  would improve the
piezoelectric response for every type of device in which currently AlN is being used, one may wonder
why not alloying with other trivalent metals, such as Y or Sc. In particular, Sc has been shown
to significantly increase the piezoelectric modulus as well.\cite{tasnadi2010origin}
Even though Sc-AlN has more exciting properties\cite{caro2015piezoelectric,tasnadi2010origin} than Cr-AlN,
the reason why ScN alloys have not taken over the resonator market so far is that the
outstanding enhancements in piezoelectric properties occur at very
high Sc concentration (Fig. 2, $x>55$\%), at which the stability
of the wurtzite phase is rather poor.
Cr-AlN has a low wurtzite-to-rocksalt transition concentration,
and therefore can offer certain piezoelectric enhancements at alloying levels that are easier to stabilize
during the synthesis.

In order to ensure significant impact of Cr-AlN alloys as materials to
outperform and replace the established piezoelectrics AlN and ZnO, two avenues should be pursued
in the near future.
First,  to benefit from the $300$\% increase in $d_{33}$ at
30\% Cr content, it is not sufficient that the rocksalt phase does not form up to that Cr concentration:
we also have to avoid  the formation of  (012) and (010)-oriented grains during growth,
which would downgrade (simply through directional averaging)
the
piezoelectric enhancements associated with the (002)-oriented grains.
To that end, we envision  changing substrates so as to enable
better lattice matching with Cr-alloys with over 25\% Cr. This can effectively prevent
 the (012) and (010) textures from emerging, therefore creating the conditions to take advantage
of the large increase in $d_{33}$ reported here.
Second, future experimental efforts should measure device performance especially to
understand the additional aspect of how  Cr content in wurtzite affects the bandgap
and whether there would be deleterious leakage effects at larger Cr concentrations.
Assuming a worst case scenario, these effects can be mitigated by co-alloying with a non-metalic atomic species (e.g. boron).

Pursuing the two directions above can make Cr-AlN suitable for simultaneous optical and mechanical resonators,\cite{yale1, yale2}
which are relatively new applications that currently exploit multi-physics aspects of AlN. 
At present, the characterization of Cr-AlN for these  multifunctional applications 
that require simultaneous engineering of the photonic and acoustic band structure
is rather incipient, and only few relevant properties of the Cr-AlN alloys are known: 
for example, for a Cr concentration of about 2\%, the bandgap is virtually unchanged, while the adsorption band 
decreases from 6 to 3.5 eV.\cite{highly2} Future theoretical and experimental work 
to investigate, e.g., photoelastic effect and optical attenuation, 
is necessary in order to fully uncover the potential of Cr-doped AlN for these applications.
For now, we surmise that the technological reason for which one would replace AlN with Cr-AlN for use in multifunctional resonators is the
trade-off between the increase in vibrational amplitude and decrease in frequency: 
while low amounts of Cr may lower the frequency somewhat, the oscillation amplitude would increase
due to larger piezoelectric response. The decrease in frequency can be mitigated
by co-doping with a small trivalent element (boron), as shown for other doped AlN 
alloys.\cite{manna2017tuning} 
Last but not least, it is worth noting that doping with Cr could enable magnetic 
polarizaton of the Cr ions in the wurtzite lattice and/or of the minority carriers: 
these effects are non-existent in pure AlN, and  could be pursued for spintronic 
applications or for low-hysteresis magnets.\cite{endo2005magnetic}

The significant increase of piezoelectric modulus reported here provides significant drive
to pursue the two directions identified above, and overcome routine barriers
towards establishing Cr-AlN as a replacement for AlN with large performance enhancements.

{\it Acknowledgments.} The authors gratefully acknowledge the support of the National Science Foundation through Grant No.  DMREF-1534503.
The DFT calculations were performed using the high-performance computing facilities at Colorado School of Mines (Golden Energy Computing Organization) and at National Renewable Energy Laboratory (NREL). Synthesis and characterization facilities at NREL were supported by the US Department of Energy, Office of Science, Office of Basic Energy Sciences,
as part of the Energy Frontier Research Center "Center for Next Generation of Materials by Design: Incorporating Metastability" under contract No.
DE-AC36-08GO28308.


\begin{thebibliography}{99}
\bibitem{nomenclature} By convention, we refer to $e_{ij}$, which couple strain and dielectric
displacement (or electric field and stress), as piezoelectric coeffcients, while $d_{ij}$,
which couple strain with electric field (or stress with dielectric displacement), are
referred to as piezoelectric moduli. In both cases, we express
the individual matrix elements in standard reduced Voigt notation.
\end{thebibliography}
\end{document}